\documentclass[aps,twocolumn,prb,preprintnumbers,amsmath,amssymb,superscriptaddress]{revtex4-1}

\usepackage{graphicx}
\usepackage{bm}
\usepackage{hyperref}
\usepackage{natbib,float}

\begin{document} 

\title{Separated transport relaxation scales and interband scattering in SrRuO$_3$,  CaRuO$_3$, and Sr$_2$RuO$_4$ thin films}

\author{Youcheng Wang$^1$, H. P. Nair$^2$, N. J. Schreiber$^2$, J. P. Ruf$^3$, Bing Cheng$^1$, D. G. Schlom$^{2,4,5}$, K. M. Shen$^{3,4}$, and N. P. Armitage$^1$\\
	\medskip
	$^1$ The Institute for Quantum Matter, Department of Physics and Astronomy,
	The Johns Hopkins University, Baltimore, MD 21218 USA\\
	$^2$ Department of Materials Science and Engineering,
	Cornell University, Ithaca, New York 14853, USA\\
	$^3$ Laboratory of Atomic and Solid State Physics, Department of Physics, 
	Cornell University, Ithaca, New York 14853, USA\\
	$^4$ Kavli Institute at Cornell for Nanoscale Science, Ithaca, New York 14853, USA
	$^5$ Leibniz-Institut f\"ur Kristallz\"uchtung, Max-Born-Str. 2, 12489 Berlin, Germany
}
\date{\today}

\begin{abstract}

The anomalous charge transport observed in some strongly correlated metals raises questions as to the universal applicability of Landau Fermi liquid theory. The coherence temperature $T_{FL}$ for normal metals is usually taken to be the temperature below which $T^2$ is observed in the resistivity. Below this temperature, a Fermi liquid with well-defined quasiparticles is expected.  However, metallic ruthenates in the Ruddlesden-Popper family, frequently show non-Drude low-energy optical conductivity and unusual $\omega/T$ scaling, despite the frequent observation of $T^2$ dc resistivity.  Herein we report time-domain THz spectroscopy measurements of several different high-quality metallic ruthenate thin films and show that the optical conductivity can be interpreted in more conventional terms. In all materials, the conductivity has a two-Drude peak lineshape at low temperature and a crossover to a one-Drude peak lineshape at higher temperatures. The two component low-temperature conductivity is indicative of two well-separated current relaxation rates for different conduction channels. In SrRuO$_3$ and  Sr$_2$RuO$_4$, both relaxation rates scale as $T^2$, while in CaRuO$_3$ the slow relaxation rate shows $T^2$, and the fast relaxation rate generates a constant background in conductivity. We discuss three particular possibilities for the separation of rates: (a)  Strongly energy dependent inelastic scattering;  (b) an almost-conserved pseudomomentum operator that overlaps with the current, giving rise to the narrower Drude peak; (c) the presence of multiple conduction channels that undergoes a crossover to stronger interband scattering at higher temperatures.  None of these scenarios require the existence of exotic quasiparticles.   However, the interpretation in terms of multiple conduction channels in particular is consistent with the existence of multiple Fermi surfaces in these compounds and with the expected relative weakness of $\omega^2$ dependent effects in the scattering as compared to $T^2$ dependent effects in the usual Fermi liquid treatment.   The results may give insight into the possible significance of Hund's coupling in determining interband coupling in these materials.  Our results also show a route towards understanding the violation of Matthiessen's rule in this class of materials and deviations from the ``Gurzhi" scaling relations in Fermi liquids. 
\end{abstract}

\maketitle

\section{Introduction}

Transport phenomenon in strongly correlated metals is in many cases still poorly understood. In Fermi liquids at low temperatures, disorder and electron-electron scattering dominate the transport. The charge resistivity $\rho(\omega, T)$ is expected to include quadratic leading terms that go as $a T^2$ and $b \omega^2$ from two-particle scattering. Nonetheless, the relation between prefactors $a$ and $b$ in general depends on the combined effect of dimensionality~\cite{rosch2005zero}, fermiology~\cite{rosch2002optical}, and details of disorder~\cite{miranda2005disorder}.  In multiband metals, where multiple conduction channels exist, understanding how the conductivity $\sigma(\omega, T)$ and resistivity $\rho(\omega, T)$ behave are important for differentiating Fermi-liquid and non-Fermi liquid behavior. In this paper, we examine the optical conductivity and the temperature- and energy-dependent relaxation rates in the strongly correlated metallic ruthenates, at the lowest energy scales with low disorder. 

The Ruddlesden-Popper family A$_{n+1}$Ru$_n$O$_{3n+1}$ of ruthenates (where A = Sr or Ca) exhibits a variety of different phases. In single-crystal form, they can be very clean and offer a unique opportunity to examine systems with strong electronic correlations and elucidate dynamical charge transport in 4d electron systems. Among them, infinite-layer ($n = \infty$) perovskite and pseudocubic SrRuO$_3$ is a ferromagnetic (FM) metal, with a Curie temperature $T_c$ of $\sim$ 160 K in single crystals~\cite{koster2012structure}. CaRuO$_3$ has the same stoichiometry and is also orthorhombic and metallic, but has larger structural distortions due to the smaller size of the Ca ion~\cite{bensch1990structure}. It is a paramagnetic metal on the verge of becoming ferromagnetic~\cite{he2001disorder,tripathi2014ferromagnetic}. Sr$_2$RuO$_4$ is a single-layer perovskite whose normal state has been regarded as a paramagnetic quasi-2D Fermi liquid~\cite{maeno1997two}, and in which unconventional superconductivity was discovered~\cite{maeno1994superconductivity}. The superconductivity in Sr$_2$RuO$_4$ is very sensitive to disorder~\cite{mackenzie1998extremely} and uniaxial strain~\cite{hicks2014strong}. Ca$_2$RuO$_4$, however, is an antiferromagnetic (AFM) insulator which can be driven by \textit{dc} current into a semi-metallic non-equilibrium steady state~\cite{okazaki2013current}.  Quantum oscillations have been observed in the above ruthenates which are metallic~\cite{mackenzie1998observation,schneider2014low,mackenzie1996quantum}. Low-temperature $T^2$ resistivity was reported in less disordered samples of SrRuO$_3$~\cite{capogna2002sensitivity} and Sr$_2$RuO$_4$ (in-plane)~\cite{barber2018resistivity}. In CaRuO$_3$, the coherence temperature scale below which resistivity follows $T^2$ was found to be $\approx$ 1.5 K in a high-quality thin film~\cite{schneider2014low}. 




However, a major puzzle in many of these materials has been reconciling their dc transport and ac (optical) properties. For instance, SrRuO$_3$ thin films have been reported to show $T^2$ resistivity below $\sim$ 30 K~\cite{mackenzie1998observation}, but non-Drude optical conductivity at THz~\cite{dodge2000low} and infrared frequencies~\cite{kostic1998non}. Measurements of CaRuO$_3$ thin films showed Drude-like conductivity below 0.6 THz and inconsistency with Fermi-liquid behavior above it~\cite{schneider2014low}. In optical measurements of Sr$_2$RuO$_4$ thin films, a universal Fermi liquid scaling of $\omega$ and $T$ of the resistivity is reported to be obeyed below 40 K, and 36 meV (corresponding to 8.7 THz).  Deviations at higher energies were accredited to ``resilient quasiparticles~\cite{stricker2014optical}. Despite extensive efforts, a comprehensive understanding of dynamical transport at the lowest energy scales in these materials is still lacking. Can the anomalous aspects in optical conductivity be understood in conventional terms or must other effects be invoked?

Here, we present a systematic time-domain THz spectroscopy (TDTS) study of three metallic ruthenates: SrRuO$_3$, CaRuO$_3$, and Sr$_2$RuO$_4$, in the form of high-quality thin films. The crucial finding is that at low temperatures, the complex conductivity of all three materials can be modeled as a sum of two Lorentzians, and smoothly evolves into a single Lorentzian as temperature is elevated. This picture gives a way of understanding deviations from the ``Gurzhi" scaling relation~\cite{gurzhi1959mutual} between $aT^2$ and $b\omega^2$ dependencies in $\rho(\omega, T)$. The observation of two separate conduction channels (instead of a single channel with strong frequency dependent inelastic scattering) with two Drude scattering rates may be understood either through a semi-classical multiband model or the possible existence of an almost conserved pseudomomentum as a result of quasi-1D Fermi surface sheets. Both scenarios do not assume the existence of exotic quasiparticles. Our observation might be generic to strongly correlated metals, give insight into the consequences of Hund's coupling in these materials, and could help explain the seemingly anomalous optical conductivity and scaling relations in a whole host of strongly correlated metals.

\section{Methods}

 In a typical TDTS experiment, a broadband THz pulse is generated by the interaction of a near-infrared femtosecond laser pulse with a voltage-biased photoconductive antenna, which is made of  two strip-shaped Au electrodes deposited on a semiconductor substrate (e.g.  LT-GaAs).  The photocarriers are excited in the semiconductor substrate and transiently accelerated by the bias voltage before recombination, resulting in a primarily dipole radiation  pattern that is collimated and coupled to free space via a hyperhemispherical Si lens. The $\sim$1 ps long THz pulse is then focused and transmitted through the sample, whose electric field is measured in the time domain using a delay stage which changes the optical path difference of the emitter and detector (another photoconductive antenna without voltage bias) branches. A reference measurement is done on a nominally identical substrate. Dividing Fourier transforms of the two time traces gives a complex transmission $T(\omega)$ which is then used to calculate complex conductivity of the thin film $T(\omega) = \frac{(1 + n)}{1+n + \sigma(\omega) d Z_0} e^{\frac{i\omega\Delta L(n-1)}{c}}$. In this expression $n$ is the substrate index of refraction measured separately,  $d$ is the sample film thickness, $Z_0$ is the impedance of free space (377 $\Omega$), and $\Delta L$ is the thickness difference between referencing substrate and  sample substrate. We accurately determined the effective $\Delta L$ from self-consistent ``first echo" measurements of the sample and substrate at different temperatures, and interpolate the results between them. For a measurement of $\Delta L$, extended scans that includes both the transmitted pulse through the sample (or reference substrate) and the first echo pulse in the sample (or reference substrate) are measured. The division of the echo field by the transmitted field is the complex transmission function $T_{Field}(\omega)$. The ratio of the field transmissions of the sample and its reference substrate encodes information about the effective $\Delta L$ in its complex phase. Owing to the reflections at the interfaces (e.g., between sample and vacuum), additional phases need to be accounted for which can be calculated through Fresnel's equations. After subtracting these additional factors, the remaining phase is associated with $\Delta L$ as $ \Delta\Phi =2 \omega \Delta L n/c$. Therefore,  $\Delta_L$ can be obtained from linear fitting of $\Delta\Phi$ as a function of frequency $\omega$. Such a process is iterative and needs to be solved in a self-consistent manner, since impedance of the sample depends on $\Delta L$ and affects the reflection coefficients at interfaces. Details about this approach can be found in the Supplemental Materials of Ref. \onlinecite{wang2020sub}. 

The SrRuO$_3$ films were grown on single-crystal (110) DyScO$_3$ substrates by molecular-beam epitaxy (MBE) in a Veeco GEN10 system. The [$\overline{1}10$] and [$001$] directions refer to the axes defined by the GdFeO$_3$-type crystal structures of the DyScO$_3$ and NdGaO$_3$ substrates. They are both indexed in the $Pbnm$ setting of space group \#62. The thickness of the film was 23 nm (SrRuO$_3$). The  adsorption-controlled growth conditions are used in which an excess flux of elemental ruthenium is supplied to the growing film and thermodynamics controls its incorporation through the desorption of volatile RuO$_x$~\cite{nair2018synthesis}. This growth regime minimizes ruthenium vacancies in the films. Ruthenium vacancies in SrRuO$_3$ films can be detected in Raman measurements and produce signatures in transport that look like the topological Hall effect, but are not~\cite{kim2020inhomogeneous, miao2020strain}. In contrast, samples with minimized ruthenium vacancies exhibit a high residual resistivity ratio (RRR), $\rho(300$K$)/ \rho (4$K$)$, of $\sim$ 70 in transport measurements, including the sample studied in this work~\cite{nair2018synthesis}. Compared to previously measured films by optical spectroscopy which only had RRRs $\sim$ 10, this film has fewer defects related to cation non-stoichiometry.  The strong dependence of spectral features on sample quality highlights the necessity for studies such as the present one utilizing SrRuO$_3$ grown by oxide MBE, which produces higher quality  SrRuO$_3$ films than those grown by pulsed laser deposition or sputtering~\cite{kacedon1997magnetoresistance, kan2013epitaxial, nair2018synthesis, macmanus2020new}. 

The  CaRuO$_3$ and Sr$_2$RuO$_4$ thin films under study were grown on 1 cm$\times$1 cm$\times$1mm single crystal NdGaO$_3$ (110) substrates in the same Veeco GEN10 system to thicknesses of 44 nm and 18.5 nm, respectively. The RRR for the CaRuO$_3$ is $\approx$ 29 along the [$\overline{1}10$] and 60 along the [001] direction. For  Sr$_2$RuO$_4$, the film thickness was chosen so as to have measurable THz transmission (2.5\%) at low temperatures, but also sufficiently high RRR $\approx$ 53. Although we could not take optical measurements below 1.3 K, it is important to point out that the disorder in the Sr$_2$RuO$_4$ film is sufficiently low such that it is superconducting~\cite{krockenberger2010growth,uchida2017molecular,nair2018demystifying}.   
\begin{figure}
	\begin{center}
		\includegraphics[width=.8\columnwidth]{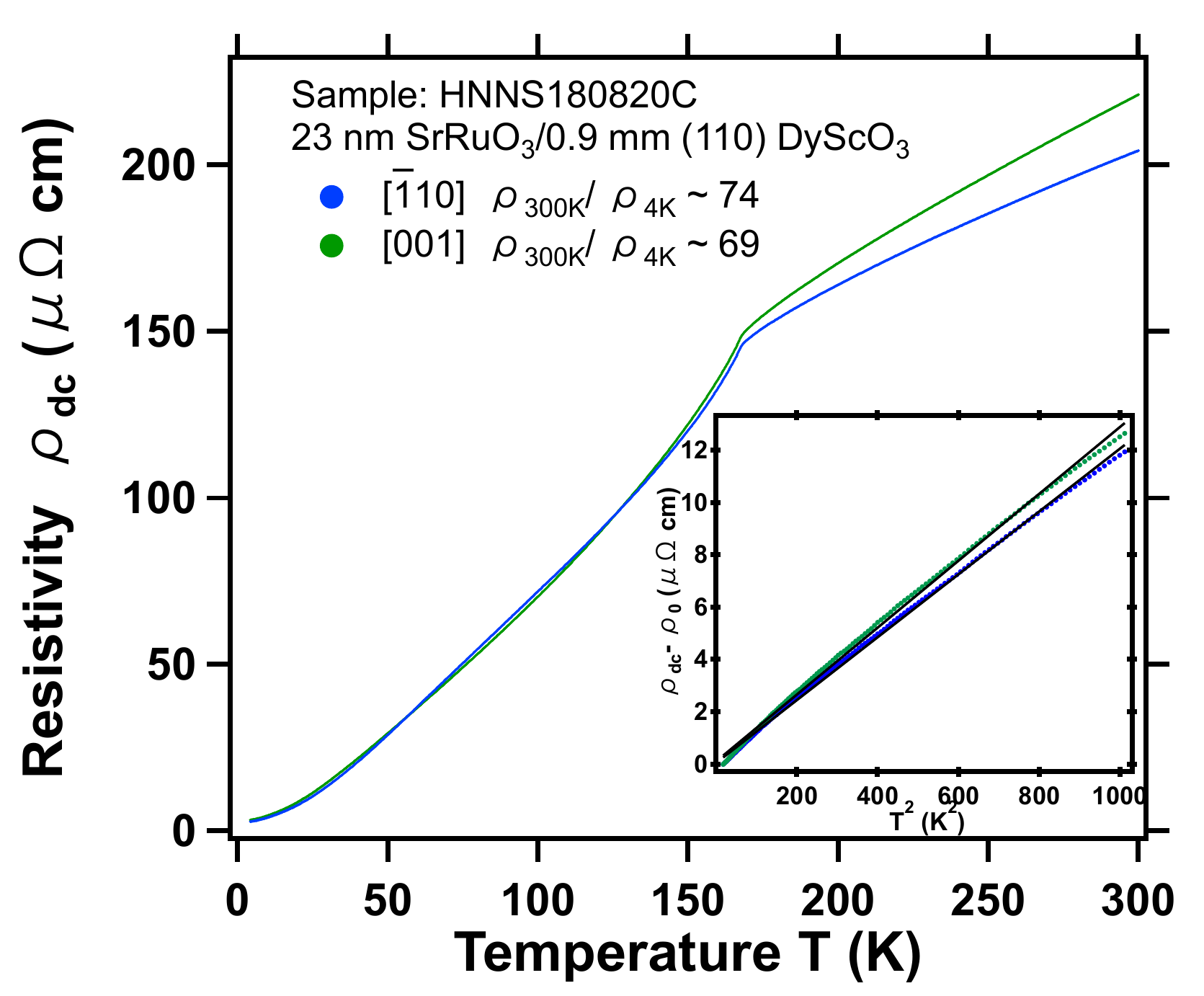}
		\centering
		\caption{(color online) dc resistivity as a function of temperature for the SrRuO$_3$ film for electric field along [$\overline{1}$10] direction.  Inset: Resistivity minus residual resistivity as a function of $T^{2}$ up until 1000 K$^2$.  The black dashed lines is the linear fit. This figure is adapted from Ref. \onlinecite{wang2020sub}. }
		\label{fig:resistivitySr113}
	\end{center}
\end{figure}


\begin{figure*}
	\begin{center}
		\includegraphics[width=1\textwidth]{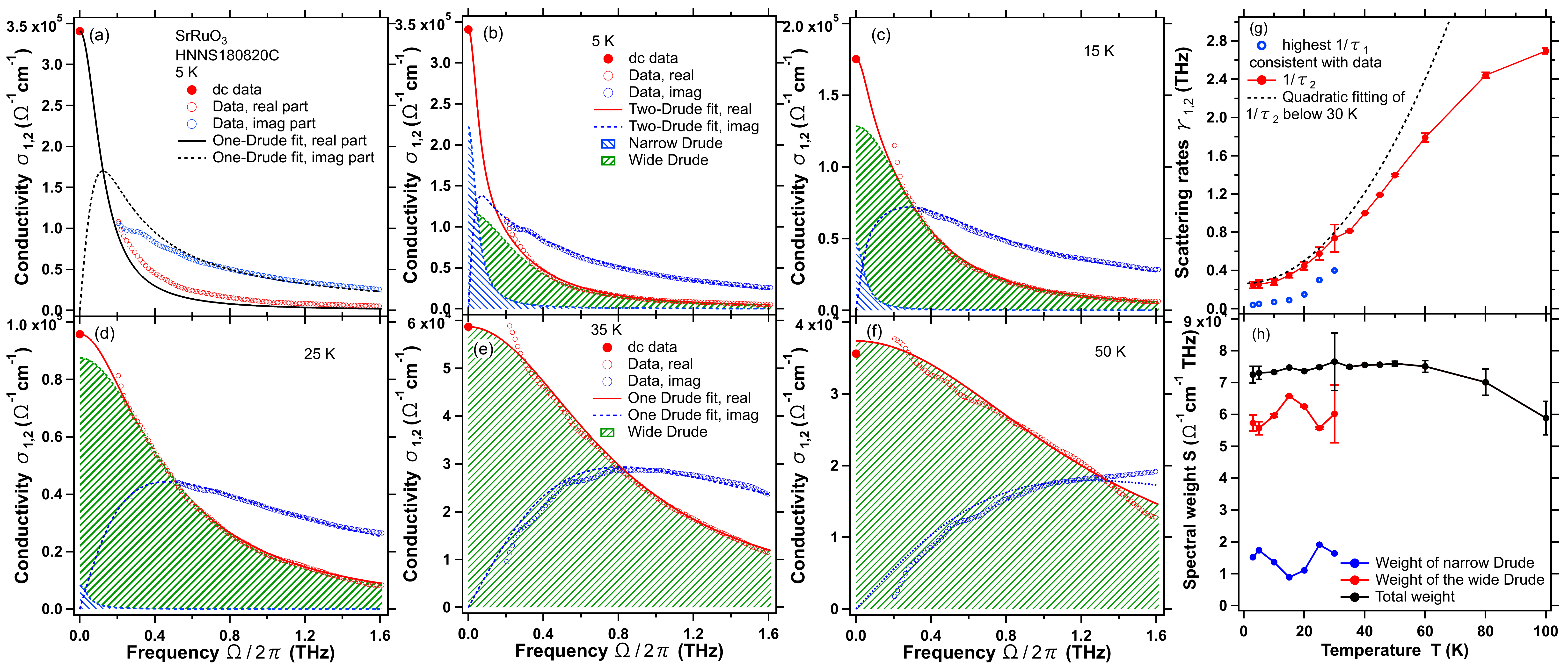}
		\centering
		\caption{(color online) (a-b) 5 K conductivity of a SrRuO$_3$ thin film sample for E//[$\overline{1}$10]. (a) Real and imaginary part of the complex conductivity, plotted with dc conductivity and a single-Drude fitting. (b) The same data as in (a) with two-Drude fitting, which is improved compared to that in (a).  (c-d) Complex conductivity at 15 K and 25 K, plotted with two-Drude fitting sharing the same legend as in (b). (e-f) Complex conductivity at 35 K and 60 K, which can be fit with a single Drude term. (g) Scattering rates for the two phenomenological Drude terms plotted against temperature. The narrow Drude term is fitted with the highest scattering rate that is compatible with the data.  (h) The spectral weights of the two Drude terms where they can be resolved separately and the total spectral weight. This figure is adapted from Ref. \onlinecite{wang2020sub}.} 
		\label{fig:srocond}
	\end{center}
\end{figure*}

\section{S\lowercase{r}R\lowercase{u}O$_3$}
SrRuO$_3$ is a 4$d$ transition metal oxide (TMO) itinerant ferromagnet.  Its perovskite structure consists of a network of corner-sharing RuO$_6$ octahedra and has a distorted orthorhombic ($Pbnm$) structure below 820 K~\cite{kennedy1998high}. The structural distortion can be described in terms of tilting and rotation of the  RuO$_6$ octahedra which results in a smaller Ru-O-Ru bond angle and electronic bandwidth~\cite{rondinelli2008electronic}. In the bulk, the Curie temperature is $T_c \approx$ 160 K and the measured saturated spontaneous magnetic moment is $\approx$ 1.1-1.6 $\mu_B$ per Ru ion~\cite{grutter2010enhanced}. In epitaxial films, the Curie temperature increases with tensile strain and decreases with compressive strain~\cite{dirsyte2011impact}. The ferromagnetism is believed to have both itinerant and localized character~\cite{koster2012structure}.  The anomalous Hall resistivity in SrRuO$_3$ changes sign from positive to negative at around 130 K and is proportional to the longitudinal resistivity squared at low temperatures~\cite{haham2011scaling}, which has been proposed to be in part ascribable to magnetic monopoles of the Berry curvature (e.g., Weyl nodes)~\cite{fang2003anomalous}. The electronic and magnetic properties of SrRuO$_3$ are difficult to be reproduced by theoretical modeling, owing to the significance of several energy scales, including Coulomb interaction, spin-orbit coupling, and Hund's coupling~\cite{dang2015electronic}.

One of the puzzles surrounding SrRuO$_3$ was its charge transport and its relation with low-energy optical conductivity. While the dc resistivity generally shows $T^2$ dependence below 30 K in low-disorder samples (e.g., Fig. \ref{fig:resistivitySr113}), previous infrared reflectivity measurements of thick films grown on SrTiO$_3$ suggest that the real part of the conductivity has an asymptotic fractional power relation with frequency $\sigma_1 \propto \omega^{-\alpha}$ where $\alpha \approx 0.5$, from 50 cm$^{-1}$ to 10,000 cm$^{-1}$ (1.5 THz to 300 THz), at 40 K~\cite{kostic1998non}. This relation is non-Drude ($\alpha$ = 2 for the high-frequency tail) and has been compared to the observations in cuprate superconductors where $\alpha \approx$ 0.7~\cite{schlesinger1990superconducting}. TDTS measurements (combined with infrared reflectivity) of thin films whose RRR is around 10 found the conductivity to be non-Drude and could be fitted with a fractional power form  \begin{equation}
   \sigma(\omega) =  \frac{A}{(1/\tau-i \omega)^{\alpha}},
	\label{frac}
\end{equation} that has also been discussed for high-temperature cuprate superconductors~\cite{schlesinger1990superconducting,el1994infrared}.  $A$ is a scale parameter and $\alpha$ was found to be $\approx$ 0.4 for a wide frequency range from 6 cm$^{-1}$ to 2,400 cm$^{-1}$ (0.18 THz to 72 THz)~\cite{dodge2000low}.  Given the sensitivity of the ground state of ruthenates to disorder, examining samples with lower residual resistivity is crucial to understand whether the value $\alpha$ strongly depends on the level of disorder. A question is then, whether there is a crossover from Fermi liquid behavior to anomalous behavior as a function of disorder.  

TDTS measurements were performed on a SrRuO$_3$ thin film grown on a DyScO$_3$ substrate. The film has $\sim $2 $ \mu\Omega\cdot$cm residual resistivity and a RRR $\approx 74$~\cite{wang2020sub}. As discussed in Ref. \onlinecite{wang2020sub}, the form of the conductivity differs at low temperature (from 3 K to $\approx$ 30 K) from that of simple Lorentzian. As shown in Fig. \ref{fig:srocond}(a), one cannot fit both the real and imaginary conductivities simultaneously with a single Drude oscillator.  As shown in Fig. \ref{fig:srocond}(b), one can fit both real and imaginary parts with two Drude oscillators.  The model we used for fitting is \begin{equation}
	\label{twoDrude}
	\sigma(\omega) = \epsilon_0 (\frac{\omega_{p_1}^2}{1/\tau_1-i\omega}+\frac{\omega_{p_2}^2}{1/\tau_2-i\omega}) -i\epsilon_0(\epsilon_{\infty}-1)\omega,
\end{equation} in which $\epsilon_0$ is vacuum permittivity, $\omega_{p_{1,2}}$ are the plasma frequencies of the two Drude terms, and $1/\tau_{1,2}$ are the corresponding scattering rates. The $\epsilon_{\infty}$ term $-i\epsilon_0(\epsilon_{\infty}-1)\omega$ accounts for the effect of high energy absorptions on the low-frequency dielectric constant. The spectral weights (Fig. \ref{fig:srocond}(h)) $A_{1,2} = \pi^2\epsilon_0 \omega_{p_{1,2}}^2$ are proportional to the amount of charge ($n_{1,2}$) from each Drude term, by the $f$-sum rule as $A_{1,2} = \frac{\pi e^2 n_{1,2}}{2 m_e}$. 

As we discussed in Ref. \onlinecite{wang2020sub}, we believe that this is a reasonable parameterization of the data due to the multiband nature of the material. The scattering rate of the narrower Drude peak is on the edge of our measurement frequency range $<$ 0.2 THz. Therefore, the plot of the small scattering rate in Fig. \ref{fig:srocond}(h) should be taken as an upper bound.  There is a smooth crossover, from two-Drude conductivity (below 30 K) to approximately single-Drude conductivity above 30 K.  It is seen that below 25 K, both scattering rates follow $T^2$ dependence. The spectral weight, defined as the area under $\sigma_1(\omega)$ of the two Drude terms, which is proportional to the number of charge carriers in the semi-classical Drude model, does not depend strongly on temperature below 30 K, while the total spectral weight is almost conserved.


The premise of multiple bands at the Fermi surface in the metallic ruthenates is supported by calculations and experiment. Band structure calculations show that in the ferromagnetic state, for the majority spins, three t$_{2g}$ bands form hole pockets at R points, while the two e$_g$ bands form an electron pocket and an electron-like Fermi surface around the $\Gamma$ point~\cite{santi1997calculation}. The distortion from the cubic structure in general results in multiple band crossings and foldings in different regions in $k$-space, as evidenced by angle resolved photoemission (ARPES) measurements of thin films~\cite{shai2013quasiparticle}.    In the Supplemental Materials of  Ref. \onlinecite{wang2020sub}, we analyzed the semi-classical equations of motion for a two-band metal with parabolic dispersions with a temperature dependent term that represents electron-electron interband scattering~\cite{gantmakher1987carrier, maslov2016optical}. Through decoupling the two equations for the velocities, we showed that the optical conductivity  for such a two-band model can always be represented as the sum of two Lorentzian terms and hence can be represented phenomenlogically, by a two-Drude expression. In a simulation, we show that the conductivity calculated from the formula derived by Maslov and Chubukov~\cite{maslov2016optical}, can be fitted with two Drude terms, no matter the choice of scattering rates within each band, or the interband scattering rates.  We will come back to the origin of the multiple peaks below.

\section{C\lowercase{a}R\lowercase{u}O$_3$}

\begin{figure}
	\begin{center}
		\includegraphics[width=.8\columnwidth]{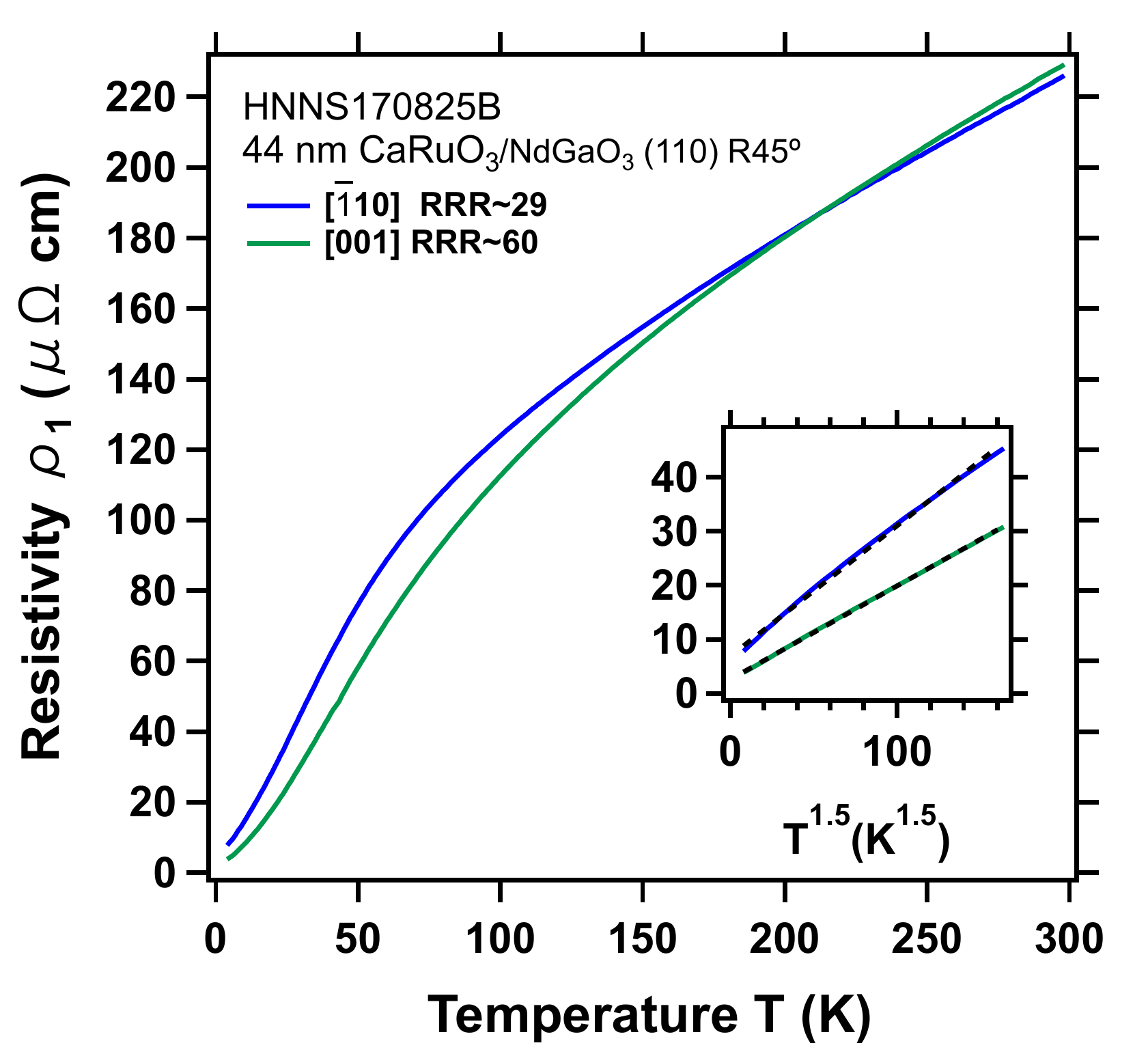}
		\centering
		\caption{(color online) dc resistivity as a function of temperature for the CaRuO$_3$ film for the two orthogonal crystallographic directions.  Inset: Resistivity minus residual resistivity as a function of $T^{1.5}$. $T^{1.5}$ fits to the data in the temperature range of 4-30 K are shown as black dashed lines.}
		\label{fig:resistivity}
	\end{center}
\end{figure}

\begin{figure*}
	\begin{center}
		\includegraphics[width=1\textwidth]{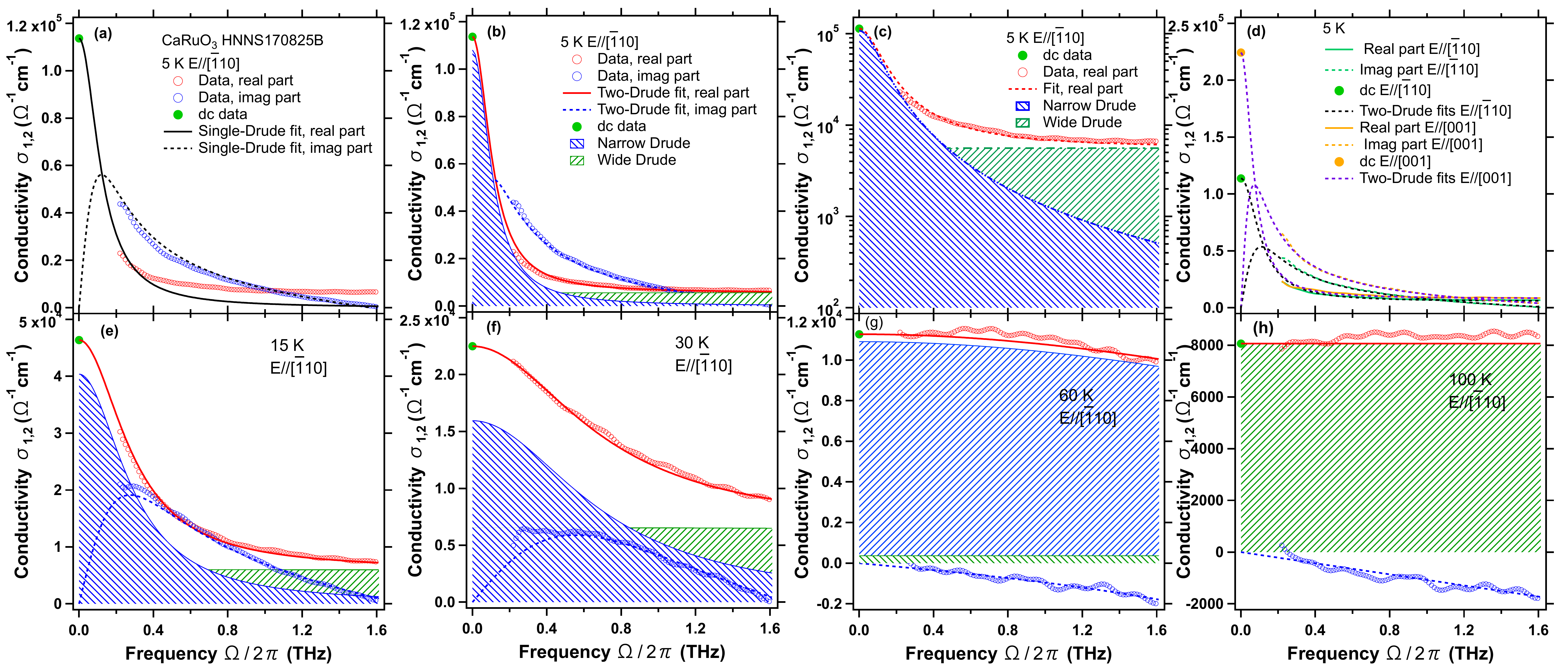}
		\centering
		\caption{(color online)(a) Real and imaginary THz conductivity of the CaRuO$_3$ thin film with dc values of 5 K data. The black dashed lines show a single-Drude term modeling. (b) The same data as in (a), but plotted with a two-Drude term fitting. The wide Drude term is fixed as a constant, since its scattering rate $\gamma_2 \gg$ measured range). (c) The same 5 K data as (a) plotted with logarithmic vertical axis. The two Drude terms in (a) are separately plotted (as indicated by blue and green shaded areas) to show how they contribute to conductivity. (d) Real and imaginary conductivity at 5 K for two orthogonal crystal axes and the corresponding two-Drude fitting.  (e-g) Real and imaginary parts of the complex conductivity with dc values for various temperatures as in the annotation of each graph, plotted with two-Drude term fitting. (h) Single Drude fitting for the data at 100 K.  Above 60 K, the conductivity is flat in the measurement range. }
		\label{fig:crocond}
	\end{center}
\end{figure*}

CaRuO$_3$ is an infinite-layer perovskite in the Ruddlesden Popper family A$_{n+1}$B$_n$O$_{3n+1}$ of metallic ruthenates. As compared to SrRuO$_3$, CaRuO$_3$ has larger GdFeO$_3$-type distortions~\cite{cao1997thermal}, slightly smaller bandwidth~\cite{goodenough1967narrow}, is paramagnetic (although can be made ferromagnetic by disorder~\cite{he2001disorder} or tensile strain~\cite{zayak2008manipulating,tripathi2014ferromagnetic}), and does not show $T^2$ resistivity until below $\sim$ 1.5 K~\cite{capogna2002sensitivity}~\cite{schneider2014low}. This low coherence temperature has motivated both experimental and theoretical investigation of the possibility of a non-Fermi liquid which contravene the Landau quasiparticle paradigm~\cite{klein1999possible,lee2002non,maiti2007observation,laad2008orbital,cao2008non,dang2015band,dang2015electronic,deng2016transport,schneider2014low,geiger2016terahertz}. A paramagnetic to ferromagnetic quantum phase transition can be induced at $x=\sim$ 0.2 in Sr$_x$Ca$_{1-x}$O$_3$~\cite{cao1997thermal}~\cite{yoshimura199917}, and therefore CaRuO$_3$ is considered to be borderline ferromagnetic and prone to magnetic fluctuations. Shubnikov de Haas oscillations~\cite{schneider2014low} and a well-defined sharp quasiparticle dispersion in photoemission~\cite{liu2018revealing} were observed in high quality thin films, but optical conductivity again hints at deviation from the Drude form and Fermi liquid predictions at high frequencies~\cite{lee2002non}~\cite{schneider2014low}. In this work, we examined complex conductivity of a high-quality (RRR $\sim$ 29) 44 nm thin film grown on a NdGaO$_3$(110) substrate by MBE. The measurement was done in two orthogonal directions, [$\overline{1}$10] and [001].

dc resistivity measured with van der Pauw geometry and analyzed by the Montgomery method~\cite{montgomery1971method} is shown in Fig. \ref{fig:resistivity}. The low-temperature part is fit to $T^{1.5}$ below 30 K for both [$\overline{1}$10] and [001] directions on the film. This smaller than 2 exponent is unusual for a clean metal, but has been seen before in CaRuO$_3$ and regarded as an indicator of non-Fermi liquid physics. The anisotropy of resistivity between the [$\overline{1}$10] and [001] directions is small at low temperatures and close to that at room temperature, but larger ($\sim$ 20\%) at intermediate temperatures. For ac conductivity in the $\sim$ THz frequency range, we do not observe any low-frequency interband transitions that might arise as a result of band gap openings due to small distortion/tilts of the RuO$_6$ octahedra (Fig. \ref{fig:crocond}). DFT+DMFT studies of ruthenates point out the importance of Hund's coupling ($\sim$ 1 eV)~\cite{deng2016transport} but corrections are usually more prominent at higher energies (or temperatures) where $\omega^{-1/2}$ is observed in infrared conductivity~\cite{lee2002non}. In previous generations of lower RRR ($\sim$ 10) samples, suppression of the far-infrared conductivity below $\omega \sim$ 2 THz is observed at elevated temperatures ($>$100 K)~\cite{lee2002non}. This gives a finite frequency peak that might originate from disorder or the substrate referencing.  However, we see no evidence for it in our present data for either crystallographic direction in our low-disorder films (Fig. \ref{fig:crocond}).

In our data, we see only a zero frequency peak for the entire range of temperatures measured (5-300 K). Figs. \ref{fig:crocond}(a), and \ref{fig:crocond}(e)-\ref{fig:crocond}(h) show real and imaginary parts of the complex conductivity at representative temperatures, with polarization of light along the [$\overline{1}10$] direction. The data with E//[001] is similar. Similar to SrRuO$_3$, the data cannot be fit with a single Drude form at low temperatures, but needs at least two Drude terms and a finite $\epsilon_\infty$  (see Eq. \ref{twoDrude}). One of the Drude terms is narrow having a $\sim $0.1 THz scattering rate $1/\tau_1$ at 5 K, and the other is wide with scattering rate $1/\tau_2$ beyond the frequency range of our measurement and is therefore approximated as a constant conductivity, as shown by the decomposed spectra in Fig. \ref{fig:crocond}(b). This functional dependence is similar to the THz conductivity of SrRuO$_3$ and Sr$_2$RuO$_4$ thin films of similar or higher RRR that can also be modeled by two Drude terms (see Ref. \onlinecite{wang2020sub} and Section \ref{s2ro}), but it is in CaRuO$_3$ that the separation of two time scales is most pronounced. Note that the THz conductivity of CaRuO$_3$ of similar RRR was reported previously~\cite{schneider2014low}, but the higher sensitivity of our experiments and accurate determination of $\Delta_L$ allow for a detailed quantitative examination. As the temperature is raised above 60 K, the conductivity restores the Drude shape and only a single Drude term is needed to fit the data, as shown in Figs. \ref{fig:crocond}(g) and \ref{fig:crocond}(h).

\begin{figure}
	\includegraphics[width=0.3\textwidth]{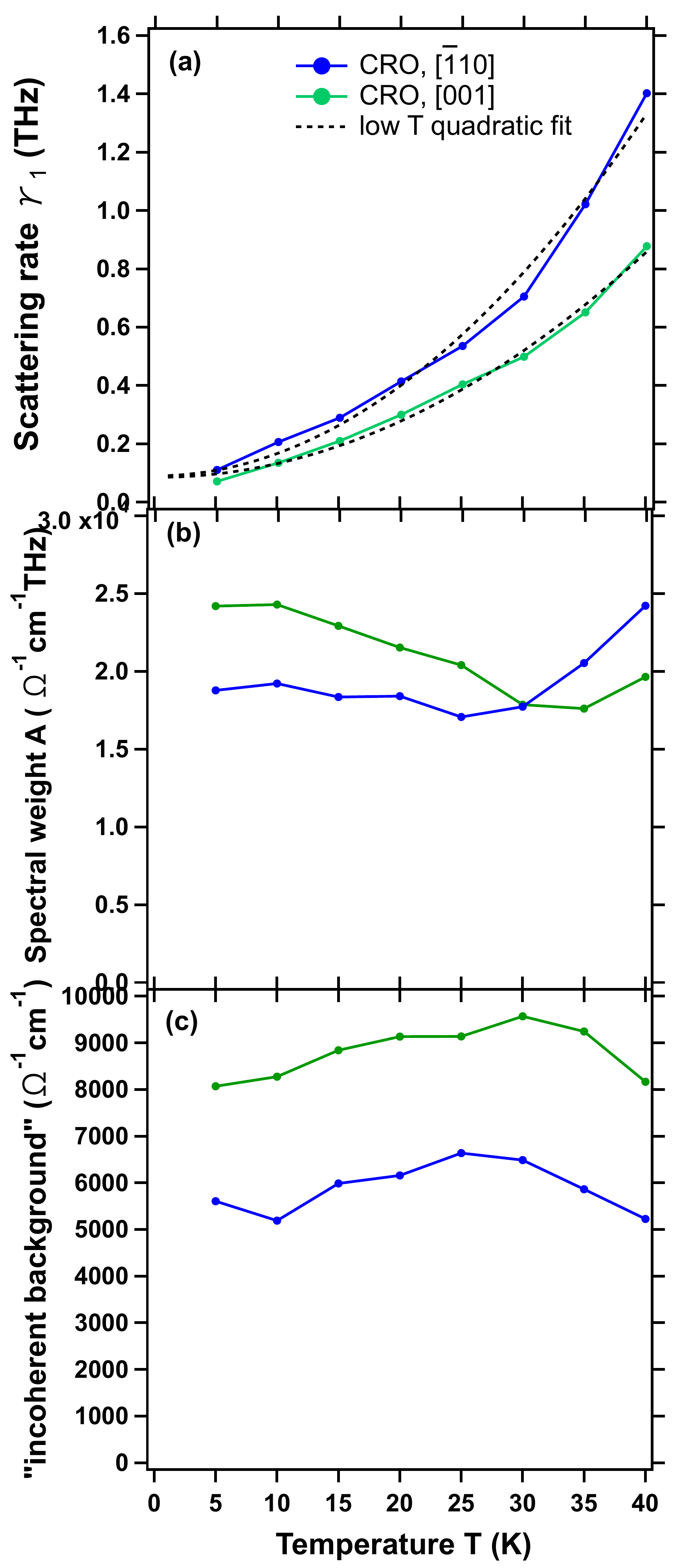}
	\centering
	\caption{(color online)(a) Scattering rates of the narrow Drude peak from two Drude modeling as a function of temperature, for the two in-plane directions in the CaRuO$_3$ thin film. The dashed black lines are quadratic fits to the data below 40 K, and extended to higher temperatures. (b) Spectral weight of the narrow Drude peak as a function of temperature. (c) The wide Drude term (incoherent background) part of the conductivity against temperature.}
	\label{fig:RtWt}
\end{figure}

The narrow Drude scattering rate as obtained from two-Drude-term modeling follows $T^2$ dependence up to 40 K (Fig. \ref{fig:RtWt}(a)). This $T^2$ might have contributions from normal electron-electron scattering (assisted by disorder), umklapp scattering, and electron paramagnon scattering~\cite{mazin1997electronic}. The scattering rates of the narrow Drude for the two orthogonal directions (blue for [$\overline{1}10$] and green for [001]) have similar residual values from disorder, but differ in terms of the prefactor for the $T^2$ dependence. The difference between two pseudocubic directions is as large as 40\% at 40 K, possibly not only because of the orthorhombic structure, but also spin-orbit interaction which makes the spin fluctuation spectrum anisotropic~\cite{mukuda1999spin}. As shown in Fig. \ref{fig:RtWt}(b), the spectral weight of the narrow Drude peak is almost conserved below 40 K.
 
 
The presence of multiple conduction channels might be related to band-dependent mass renormalizations. Photoemission~\cite{liu2018revealing} and quantum oscillations~\cite{schneider2014low} as well as band theory calculations of CaRuO$_3$~\cite{liu2018revealing} show the presence of multiple small electron/hole pockets and several Fermi surface sheets across $E_F$. An \textit{in-situ}  photoemission study of CaRuO$_3$ films observed multiple heavy bands (effective mass $m^* \sim$ 13.5 $m_e$) in the 30 meV (corresponding to 7.3 THz) region close to $E_F$.  Since the wide Drude background has a scattering rate we could not determine experimentally, we could not track its spectral weight, like we did for the narrow Drude peak. Nevertheless, the trend that two components merge into a single Drude is evident (e.g., Figs.\ref{fig:crocond}(g) and \ref{fig:crocond}(h)). 

In both SrRuO$_3$ and CaRuO$_3$, the spectral weight of the narrow Drude component does not vary considerably at low temperatures (Figs. \ref{fig:srocond}(h) and \ref{fig:RtWt}(b)). However, one empirical difference between these compounds is that when the two Drude components merge, and become less distinguishable from one another (at around 30 K for SrRuO$_3$ and around 60 K for CaRuO$_3$), the narrow Drude term in SrRuO$_3$ loses spectral weight to the wide Drude (Figs. \ref{fig:srocond}(e) and \ref{fig:srocond}(f)), while the narrow Drude in CaRuO$_3$ appears to gain spectral weight (Fig. \ref{fig:crocond}(g)). 



\section{S\lowercase{r}$_2$R\lowercase{u}O$_4$}
\label{s2ro}
Sr$_2$RuO$_4$, the $n=1$ Ruddlesden-Popper phase, is isostructural to La$_2$CuO$_4$, a parent compound of cuprate high-temperature superconductors. Since the discovery of superconductivity in single crystalline Sr$_2$RuO$_4$ more than two decades ago~\cite{maeno1994superconductivity,armitage2019superconductivity}, considerable efforts were devoted to understanding the order parameter (OP) of the superconducting phase, and the origin of unconventional superconductivity in ruthenates~\cite{mackenzie2003superconductivity,mackenzie2017even}.The normal state of Sr$_2$RuO$_4$ is usually regarded as a clean Fermi liquid~\cite{capogna2002sensitivity}. Three bands give the Fermi surface, in which hole-like $\alpha$ and electron-like $\beta$ bands are quasi-1D, and the electron-like $\gamma$ band is quasi-cylindrical~\cite{oguchi1995electronic}.  In the superconducting state, muon spin relaxation~\cite{luke1998time} and optical polar Kerr experiments~\cite{xia2006high} showed signatures of time-reversal symmetry breaking. Nuclear magnetic resonance (NMR) seemed to suggest odd-parity OP, among which the $p_x \pm ip_y$ state was a viable candidate.  This is a state that is analogous to superfluid helium-3~\cite{rice1995sr2ruo4}.  Nevertheless, Pauli-limited upper critical fields~\cite{yonezawa2013first}, absence of edge currents~\cite{mackenzie2017even}, and recent observation of pronounced drop of NMR Knight shift across the superconducting transition seem to contradict the previously most favored $p_x \pm ip_y$ OP~\cite{pustogow2019constraints, ishida2020reduction}. The superconductivity can be enhanced through uniaxial strain which is thought to induce a Lifshitz transition, where a van Hove singularity in the $\gamma$ band crosses Fermi energy~\cite{hicks2014strong,steppke2017strong}. The superconductivity is sensitive to disorder and destroyed when the normal state resistivity is above 1.1 $\mu \Omega\cdot$cm in single crystals~\cite{mackenzie1998extremely}.  As mentioned above, the present films are among the first to be grown with sufficiently low disorder that they are routinely superconducting~\cite{krockenberger2010growth,uchida2017molecular,nair2018demystifying}.  The presence of multiple bands, spin-orbit interaction, and possibly spatial inhomogeneity allows a variety of odd- and even-parity OP candidates~\cite{ramires2019superconducting}.

Although puzzles about the superconductivity are still to be settled, the normal state has been studied by both transport and optical experiments. Shubnikov de Hass oscillations were reported~\cite{mackenzie1996quantum, ohmichi2000magnetoresistance} and Fermi liquid-like scaling of the optical conductivity were identified in the regime of $\hbar \omega \leq$ 36 meV (corresponding to 8.7 THz) and $T\leq$ 40 K~\cite{stricker2014optical}, both of which are consistent with $T^2$ resistivity at low temperatures. Takahashi et. al. reported the in-plane THz conductivity of a $\sim$ 120 nm thin film grown on a (LaAlO$_3$)$_{0.3}$(SrAl$_{0.5}$Ta$_{0.5}$O$_3$)$_{0.7}$ (LSAT) (001) substrate~\cite{takahashi2014plane}. The film had a resistivity of 2.7 $\mu \Omega$ cm at 2 K (giving a RRR $\sim$ 40) and with a resistivity that obeyed  T$^2$ scaling up to 20 K. A multiple-band Drude Lorentz model was used to fit the dc conductivity, THz conductivity (from 2-8 meV, or 0.5-1.9 THz), and Hall coefficients simultaneously. It was claimed that the scattering rates for the $\alpha$, $\beta$, and $\gamma$ bands are 0.38, 0.34, and 0.48 THz at 4 K, respectively. 

We performed TDTS measurements on a high-quality ($\rho(4K) \approx$ 2.1 $\mu \Omega\cdot$cm) Sr$_2$RuO$_4$ film grown on NdGaO$_3$ substrates. The low-temperature normal state conductivity is separable into two Drude terms with two independent decay rates. This two-Drude observation is qualitatively similar to our findings in clean SrRuO$_3$ ($\rho(4K) \approx$ 2.6 $\mu \Omega\cdot$cm) and CaRuO$_3$ ($\rho(4K) \approx$ 3.9 $\mu \Omega\cdot$cm) which are ferromagnetic and paramagnetic metals, respectively. So while negative violation of Matthiessen's rule was found in SrRuO$_3$ and CaRuO$_3$, it has not been reported for Sr$_2$RuO$_4$. The question remains for the normal state of Sr$_2$RuO$_4$ as to whether there are multiple channels of conduction and whether these channels can be assigned to different bands. 

\begin{figure}
\begin{center}
	\includegraphics[width=1\columnwidth]{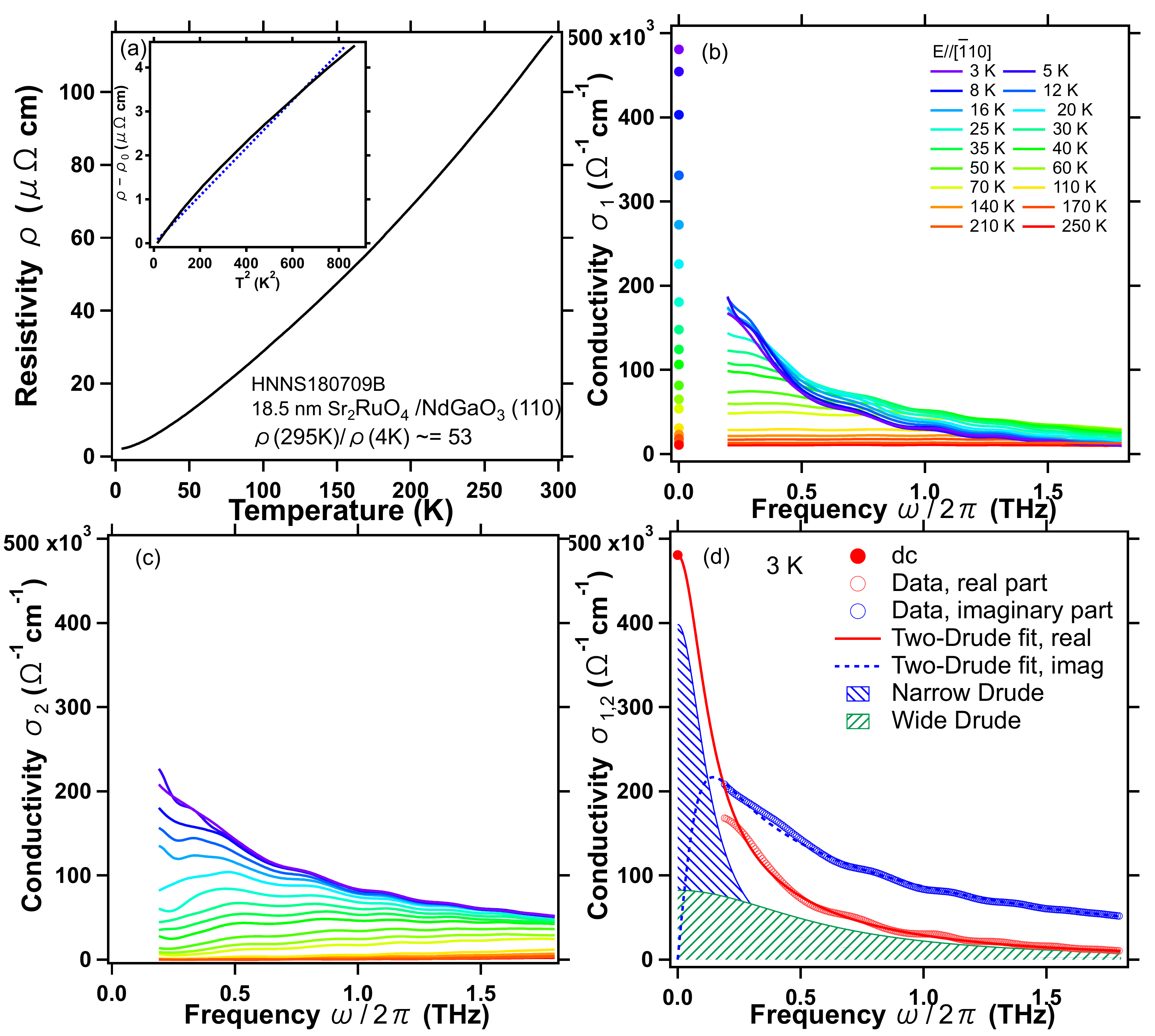}
	\centering
	\caption{(color online) (a) dc resistivity as a function of temperature for the Sr$_2$RuO$_4$ film. The data is a geometric mean of the $[\overline{1}10]$ and $[001]$ directions of the NdGO$_3$ substrate.  Inset: Resistivity minus residual resistivity as a function of temperature squared.  Fits to the data in the temperature range 2-32 K is shown as black lines. (b) Real and (c) imaginary parts of the THz conductivity $\sigma_{1,2}$ from 3 K to 250 K. The markers at zero frequency are dc conductivity. (d)  Two Drude fit of dc plus THz data at 3 K. dc conductivity, real and imaginary parts of the optical conductivity are fitted simultaneously. For all THz data, $E//[\overline{1}10]$.}
	\label{fig:fig1}
\end{center}
\end{figure}

Fig. \ref{fig:fig1}(a) shows dc resistivity measured in the van der Pauw geometry. The inset shows a tentative $T^2$ fit to the data $<$30 K which shows only slight deviations. The temperature dependence of the real and imaginary parts of the THz conductivity are shown in Figs. \ref{fig:fig1}(b) and \ref{fig:fig1}(c). At high temperatures, both parts of the conductivity do not show much frequency dependence, meaning the scattering rate exceeds the measured frequency range. As the temperature goes down, both real and imaginary parts increase, and the spectral weight is transferred into low frequencies. The imaginary part of conductivity gradually surpasses the real part, which is a signature of a narrow zero frequency conducting channel.

\begin{figure*}
	\begin{center}
		\includegraphics[width=1\textwidth]{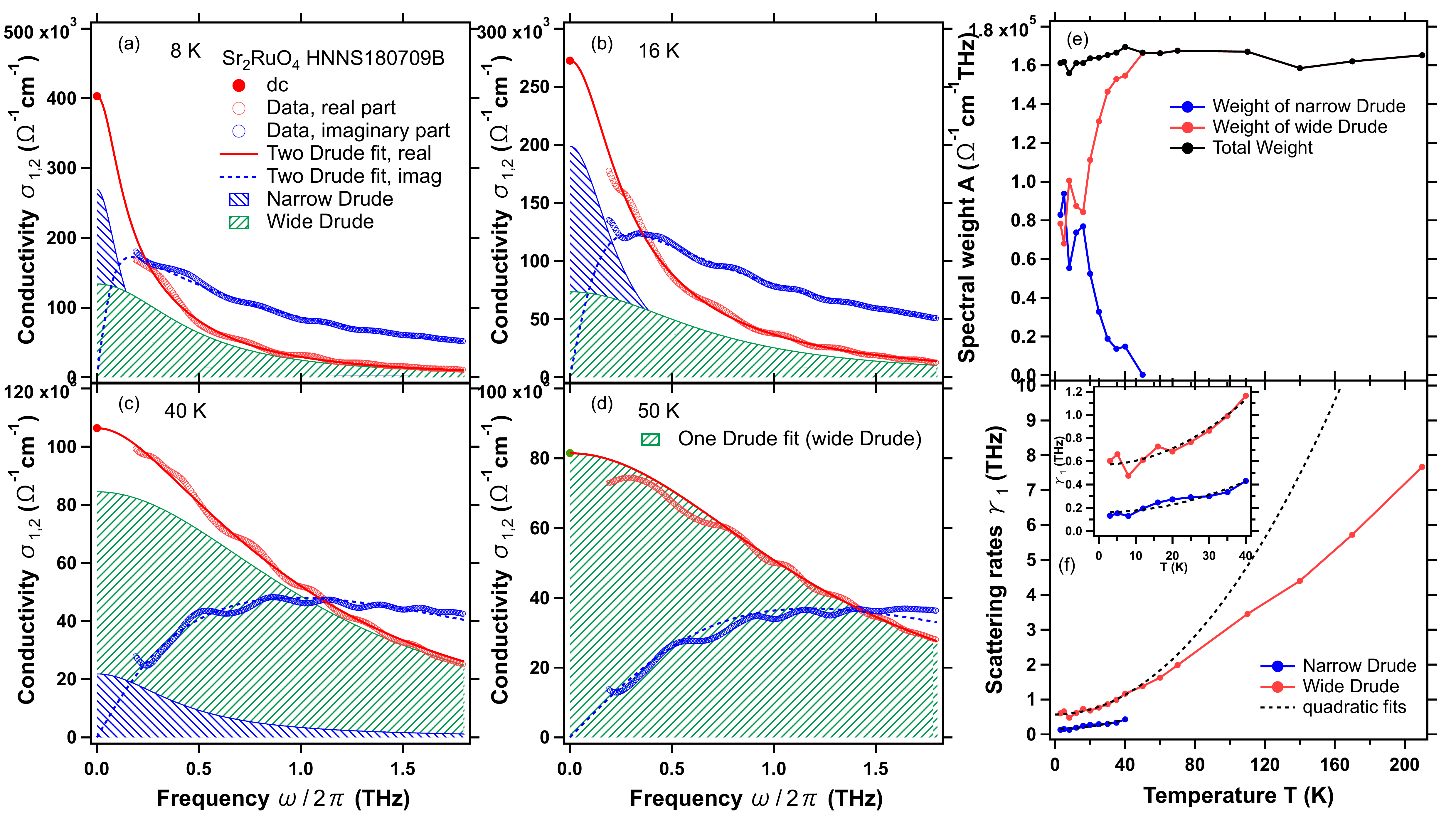}
		\centering
		\caption{(color online)(a-d) Real and imaginary THz conductivity of the Sr$_2$RuO$_4$ thin film with dc conductivity at (a)8 K, (b)16 K, (c)40 K, and (d)50 K. Fitting is presented in dashed lines. For data $\leq$ 40K, two Drude terms are required. The 50 K data can be approached by a single Drude model. Green and blue shades correspond to a narrow and wide Drude term, respectively. (e) Spectral weight and (f) scattering rates of the two Drude terms as a function of temperature. The black dashed lines in (f) are quadratic fitting.}
		\label{fig:fig2}
	\end{center}
\end{figure*}

The data below 40 K cannot be fit with a single Drude.  Analogous to SrRuO$_3$ and CaRuO$_3$, one can fit the conductivity with two Drude terms.  In Fig. \ref{fig:fig1}(d), the 3 K conductivity shows that the spectra can be decomposed into two Drude components with rather different decay rates. The narrow peak which is part of the current that decays more slowly accounts for most of the conductivity at low frequencies while the wide peak almost overlaps with the total conductivity above 1 THz.  Figs. \ref{fig:fig2}(a)-\ref{fig:fig2}(d) shows the temperature evolution of the two Drude components using the data at a few representative temperatures. The conductivity spectra are well fit with two Drude components for 8 K, 16 K, and 40 K, while the conductivity at 50 K is fitted with a single Drude term. The scattering rates of both components increase with temperature (Fig. \ref{fig:fig2}(f)). Both follow $T^2$ dependence below 40 K. From fitting, the ratio of the residual scattering rate $\gamma^w_0/\gamma^n_0 = 0.57/0.16 \approx$ 3.5, and the ratio of the coefficients for $A T^2$ dependence is $A_w/A_n \approx$ 2.1.  The spectral weights are plotted in Fig. \ref{fig:fig2}(e). The two Drude components have similar spectral weight at low temperatures up to 20 K and as the temperature is increased to 50 K, all the spectral weight is transferred to the wide Drude component. The total spectral weight is almost unchanged from 3 K to 200 K.  As noted above, Takahashi et. al. reported the in-plane THz conductivity of a low disorder $\sim$ 120 nm thin film grown (RRR $\sim$ 40) on a (LaAlO$_3$)$_{0.3}$(SrAl$_{0.5}$Ta$_{0.5}$O$_3$)$_{0.7}$ (LSAT) (001) substrate~\cite{takahashi2014plane}.  With a combination of dc conductivity, Hall resistivity, and THz conductivity (from 2-8 meV, or 0.5-1.9 THz), they fit to a three Drude model and found scattering rates at low temperature of 0.38, 0.34, and 0.48 THz (which they assigned to the $\alpha$, $\beta$, and $\gamma$ bands).   In contrast, we used a more minimalist model and have no need for three Drudes when two will do.   In this regard the choice of three Drudes was probably motivated by existence of the three known bands of the band structure and not by a necessity for the fits as two of the three scattering rates were found to be similar.

\section{Discussion}

How to understand these results?   In a typical case for a single band metal at low temperature the frequency dependent conductivity assumes a Lorentzian form, which indicates the current-current correlation decays exponentially in time.  This corresponds to frequency independent scattering, which occurs in the typical case when the current decay is dominated by scattering from static impurities, and quantum statistics are unimportant~\cite{singh2016electronic}. The robustness of the Drude conductivity relies on the law of large numbers and self-averaging such that many separate scattering events average to a single effective scattering rate that can be applied to all electrons.  Although a single Lorentzian Drude lineshape will be realized in the simplest cases, there are a number of reasons where deviations from a simple Lorentzian may be observed.   Inelastic frequency dependent scattering can be significant, there can be geometric cancelations of the scattering rate, or multiple relatively independent conduction channels can exist.  We will consider all these possibilities below.

Although disorder scattering is dominant at low temperature and frequency, inelastic scattering will occur at finite temperature and frequency in all real materials.   The question is its relative size.  Within Matthiessen's rule, the scattering rate of a conducting channel will be the sum of elastic and inelastic scattering rates.   Inelastic scattering is dominated by electron-electron scattering at the lowest temperatures and generally goes as $\omega^2$ and $T^2$.  However how this scattering manifests in the transport depends on details.  Conventional electron-electron scattering of charge in a single parabolic band does not contribute to the resistivity as the momentum-conserving scattering is also velocity-conserving.  However, intraband umklapp scattering does decay the crystal momentum (and hence the current), and gives rise to a $T^2$ scattering rate~\cite{landau1937properties,baber1937contribution}.  Moreover, interband electron-electron scattering such as present in a multi-band metal conserves crystal momentum but can still degrade the current~\cite{gantmakher1987carrier}. A non-vanishing matrix element for $T^2$ scattering is constrained by momentum and energy conservation and so whether or not the $T^2$ term is observed is shown to depend on the dimension, topology, and shape of the Fermi surfaces~\cite{pal2012resistivity}.

\begin{figure}
	\includegraphics[width=0.5\textwidth]{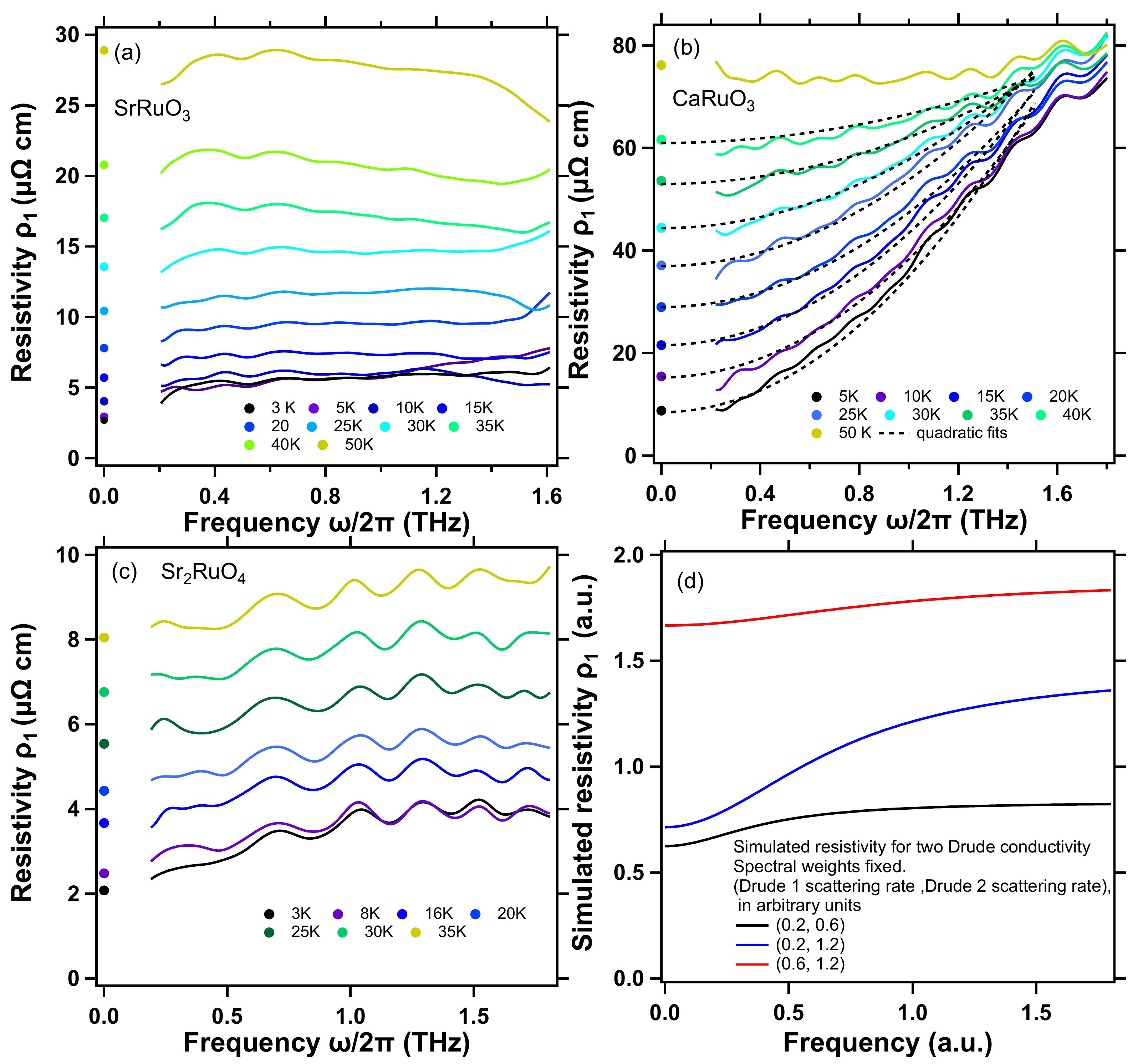}
	\centering
	\caption{(color online) (a-c) Real part of the complex THz resistivity $\rho_1(\omega)$ at different temperatures for the SrRuO$_3$, CaRuO$_3$, and SrRuO$_3$ thin films, respectively. E// [$\overline{1}10$] for all the three samples. Quadratic fitting in (b) is presented with black dashed lines. (d) Simulated $\rho_1(\omega)$ by inverting the two-Drude conductivities. The scattering rates $\gamma_1$ and $\gamma_2$ (in a.u.) for the two Drude terms are (0.2, 0.6), (0.2, 1.2), and (0.6, 1.2), for the three modeled curves, black, blue, and red, respectively. The spectral weight for the two Drude terms are set to be equal and identical for the three situations.}
	\label{fig:ResAll}
\end{figure}

Additional insight can come from looking at the frequency dependence the complex dynamical resistivity $\rho_1(\omega)$. Figures \ref{fig:ResAll} (a)-\ref{fig:ResAll}(c) show the real part of $\rho(\omega)$ of all three materials, calculated by inverting the complex conductivity, \begin{equation}
\label{complexresistivity}
\rho(\omega) = (\frac{1}{\sigma_1 + i \sigma_2- i \epsilon_0(1-\epsilon_{\infty})\omega})^*, 
\end{equation} where ``*" indicates complex conjugation. The $\epsilon_{\infty}$ term is subtracted before the inversion because it arises from higher energy excitations that make their effects felt at lower frequency only through polarization. The dc values are plotted along with the frequency-dependent resistivity.   A strong frequency dependence is seen in all cases, although there are both commonalities and differences between the different materials. The resistivity of SrRuO$_3$ and Sr$_2$RuO$_4$ at low temperatures and low frequencies show weaker dependencies.   They also both show a small mismatch with the dc values, perhaps owing to the existence of the narrow Drude peak.  CaRuO$_3$ shows the strongest frequency dependence in the frequency range (0.2-1.6 THz) and can be fitted well with an $\omega^2$ dependence up to 1.5 THz (the dashed lines in Fig. \ref{fig:ResAll}(c)).  In all cases as temperature increases, the frequency dependence of the resistivity becomes weaker, which corresponds to the behavior of a system with a single Drude peak. 

The extended Drude model has been applied to many correlated metals that are expected to have a single conduction channel~\cite{maslov2016optical,allen1971electron,allen1977optical,armitage2009electrodynamics}.   It introduces a phenomenological frequency dependent scattering rate and mass that accounts for the effects of inelastic energy dependent scattering.   These parameters are proportional to the real and imaginary parts of the frequency dependent complex resistivity respectively.    However, a question arises in the current case whether or not the frequency dependent resistivity is indicative of strong inelastic scattering, or instead arises from the superposition of multiple conduction channels that themselves have little frequency dependence.  In principle the extended Drude analysis is only valid when only a single conduction channel exists.   We will come back to this point below.  However for a single band metal, a conventional explanation for the strong $\omega^2$ resistivity seen most clearly in CaRuO$_3$ is related to umklapp scattering.  One can analyze the inelastic part of the scattering for a single-band metal in the limit of $  \omega \gg \gamma(\omega) $. It has been shown that for 3D or 2D Fermi surfaces, in the presence of umklapp scattering, the conductivity is approximately a constant (the incoherent background) $\sigma(\omega) \sim \frac{U^2 k_F^5}{v_F^4}$, and the scattering rate $\gamma(\omega) \propto \omega^2$~\cite{rosch2005zero, rosch2006optical}.  Here $U$ is the screened interaction potential, $k_F$ and $v_F$ are the Fermi wavevector and Fermi velocity, respectively.  For small Fermi surfaces in 3D, where umklapp scattering is suppressed, the conductivity is still approximately a constant but smaller by a factor 
$(k_F a)^4$, where $a$ is the lattice constant. The relations do not hold for small 2D Fermi surfaces where umklapp scattering is absent. In that situation, $\sigma(\omega) \propto \omega^2$ and $\gamma(\omega) \propto \omega^4$. In our CaRuO$_3$ data, the high-frequency tail of the conductivity is almost a constant (e.g., Fig. \ref{fig:crocond}(b)), which could reflect strong umklapp scattering in 3D or 2D. Multi-particle electron scattering leads to a term in the temperature-dependent scattering rate that goes as $\gamma(\omega \rightarrow 0, T) \propto U^{n} T^{2n-2}$, in which the integer $n$ depends on the filling of the band $\sim G/(2k_F)$ where $G$ is the reciprocal lattice vector.~\cite{rosch2005zero, rosch2002optical}. This temperature scaling agrees with the scattering rate of the narrow Drude peak, which is approximately $\gamma(\omega \rightarrow 0, T)$ (Fig. \ref{fig:RtWt}).  Anomalously strong umklapp scattering might be a reason for low coherence scale of CaRuO$_3$. 


Another possibility for the origin of the two Drude peaks is associated with ``geometric cancellation" of the scattering rate and the presence of an almost conserved current in clean metals~\cite{rosch2006optical}. It was shown by Rosch that in 1D Fermi surface sheets, umklapp scattering does not decay a pseudomomentum which connects the two sheets with a sign function. This quantity overlaps with the velocity operator (and current) when the metal is away from commensurate filling~\cite{rosch2002optical} and therefore contributes to transport, leading to partial conservation of the current.   Although the majority of current is proportional to crystal momentum and decays by umklapp scattering, this long lived pseudomomentum leads to an additional narrow Drude peak.    There are quasi-1D Fermi surface sections in Sr$_2$RuO$_4$, but less is known about the detailed shaped in the other materials.   The role of this kind of physics should be further investigated.

 
However, we consider that by far the most likely explanation for the multiple Drude peaks is multiband transport.  In a multiband metal where different carriers each contribute a Drude-like peak, the form of the low-temperature conductivity will obviously deviate from the single Drude form in different scattering rates for the different channels exist~\cite{kuz2002manifestation,mirri2012optical}.  However, in the presence of interband scattering it is not \textit{a priori} obvious that the conductivity can be decomposed into separate Drude terms.   However, since  electron-electron interband scattering vanishes as $\omega^2$ and $T^2$ and moreover that $T^2$ scattering dominates over $\omega^2$ scattering, it is reasonable that at low temperature each conduction channel is independent and contributes an independent Drude (It was shown by Gurzhi in the 1950s that $\rho(T,\omega)$ and hence the scattering rate was proportional to $ T^2 + A(\hbar/2\pi k_B)^2\omega^2$ with a coefficient A that will be 1 due to requirements of the ``first-Matsubara-frequency rule” for boson response~\cite{gurzhi1959mutual,maslov2016optical,maslov2012first}).  However as long as the scattering is not strongly energy dependent, a two Drude decomposition appears to be even more general than that.  In Ref. \onlinecite{maslov2016optical}, the authors calculated the optical conductivity from a two-band model, with equations of motion taken from Ref. \onlinecite{gantmakher1987carrier}. We analyzed the same equations of motion in Ref. \onlinecite{wang2020sub}, and found that for two bands, even in the presence of interband electron-electron scattering, the equations of motion can always be decomposed into the equations of motion for two effective Drude models (with highly effective parameters). Simulations of the optical conductivity given by the formulae in Ref. \onlinecite{maslov2016optical} can be fit with two Drude terms for all of band parameters we have tested.  This may explain the two-Drude to one-Drude term crossover with increasing temperature (at $\sim$ 30 K to 60 K), in which the spectral weight of one of the effective Drude terms vanishes as interband electron-electron scattering increases.   This can occur in partially compensated semimetals~\cite{maldague1979electron,baber1937contribution}.  At low temperatures, scattering is dominated by elastic scattering off impurities but as temperature is raised, both intraband electronic scattering, and interband electronic scattering are enhanced. In the presence of electrons and holes, there is effectively an equation of motion for the total momentum and another equation for the relative velocities. In a compensated metal, the relative velocity vanishes with sufficiently strong electron-hole scattering, which acts as a viscous damping force to make the two band velocities equal~\cite{maldague1979electron,baber1937contribution}. As a result, the conductivity is only left with one Drude term corresponding to the total momentum of the two bands velocities.   Similar physics could be playing a role here where different bands act individually at the lowest temperature where interband electron-electron scattering is weak.  At higher temperature the interband electron-electron scattering becomes enhanced and only a single highly collective conducting feature is retained.



At low temperature, if the two channels of conduction come from two independent bands and interband scattering is small, then each channel derives from the scattering rates and density of states of each band separately.  It would be nice to understand for the multiple components observed if their features can be connected to aspects of the band structure.   One would want to understand if the relative sizes of scattering rates, spectral weight, and  coefficients of the inelastic $T^2$ scattering can be understood based on the gross aspects of the band structures and the expectations for 2D vs. 3D effects.  However, this appears challenging.  For parabolic bands, the elastic scattering rate of each band would be expected to be proportional to the density of states~\cite{zawadzki1971elastic} (proportional to mass ($m$) in 2D and $m \sqrt{ E_F^3}$ in 3D).   The spectral weight goes like $E_F$ in 2D and $ \sqrt{ m E_F^3}$ in 3D.   The coefficient of the inelastic scattering has leading proportionality of $1/E_F$ in both dimensions.  One may hope that Sr$_2$RuO$_4$ is closer to the 2D limit and SrRuO$_3$ is closer to the 3D limits, and although there is some correspondence between features, a detailed analysis does not hold up.  For 3D Fermi surfaces, the density of states is larger for larger Fermi surfaces, and therefore the impurity scattering rate would be expected to have a Drude term with larger spectral weight.    This is as observed for SrRuO$_3$   On the other hand, in contrast (as shown in Figs. \ref{fig:fig2}(e)-\ref{fig:fig2}(f)), for Sr$_2$RuO$_4$, the spectral weights of the two features at low temperatures are close to each other, but the coefficient of the inelastic $T^2$ scattering is larger for the term with the larger elastic scattering.   As spectral weight and the inelastic scattering are expected to scale inversely with each other this is against the expectation for 2D.

To understand the shape of $\rho_1(\omega)$ in a two-band metal,  we performed simulations where frequency-dependent resistivity is generated by summing up multiple Drude channels in the conductivity.   Each Drude channel has a frequency independent scattering, but as one can see in Fig. \ref{fig:ResAll}(d)) the resistivity gives the appearance of frequency dependent scattering (if it was interpreted as a single channel model).   Simulations were performed using the following parameters. The two Drudes components were chosen to have equal spectral weight for simplicity and kept identical for all three situations (black, blue, and red curves in Fig. \ref{fig:ResAll}(d)). Under these circumstances, the simulations show that the resistivity will generically be an increasing function of frequency.   When the scattering rates are close to each other, the simulated resistivity increases more slowly with frequency (the red curve). The resistivity shows the strongest frequency dependence when their scattering rates are the most different. Figure \ref{fig:ResAll}(d) demonstrates this for the case when the larger scattering rate is 6 times the smaller one (the blue curve). 

This is very general behavior.  When two Drude terms with frequency-independent scattering are added in the complex conductivity, the resulting complex resistivity shows a $\omega^2$ dependence at low $\omega$. This low-frequency dependence can be seen by Taylor expanding $\rho_1(\omega)$ at $\omega= 0$.\begin{equation}
	\label{quadraticomega} \begin{split}
	\rho_1 (\omega) = \textup{Re}(\frac{1}{\frac{\sigma_A}{1-i \omega/\gamma_A}+\frac{\sigma_B}{1-i \omega/\gamma_B}})\\
	= \frac{1}{\sigma_A + \sigma_B } + \frac{\sigma_A \sigma_B(\gamma_A-\gamma_B)^2}{\gamma_A^2\gamma_B^2 (\sigma_A + \sigma_B)^3} \omega^2 + \mathcal{O}(\omega^4)
	\end{split}
\end{equation}  In the above expression, $\sigma_A$ and $\sigma_B$ are the dc conductivities of the two Drude terms respectively. $\gamma_A$ and $\gamma_B$ are the corresponding scattering rates. The leading term is $\omega^2$. Again we emphasize that this $\omega^2$ arises not because of the energy dependence of electron-electron scattering , but due to the inversion of the superposition of two Lorentzians.


Because of the very large disparity in the two fitted scattering rates of CaRuO$_3$, a quadratic fitting of $\rho_1(\omega)$ of CaRuO$_3$ allows for the most straightforward comparison with experiments in other materials.   As discussed above, it was shown by Gurzhi in the 1950s for a single band Fermi liquid  that $\rho(T,\omega) \propto T^2 + A(\hbar/2\pi k_B)^2\omega^2$. As shown by Gurzhi and later discussed by Maslov and Chubukov~\cite{gurzhi1959mutual,maslov2016optical,maslov2012first}, for pure umklapp scattering the coefficient A has to be 1 due to requirements of the ``first-Matsubara-frequency rule” for boson response. In actuality many materials that otherwise look like Fermi-liquid like metals show notable differences of $A$ from 1~\cite{nagel2012optical, sulewski1988far, dressel2011quantum}. To investigate the applicability of this expression, we take the coefficient for $T^2$ from the fitted narrow Drude peak. From the frequency dependent resistivity fits, the value of A is estimated to be 2.8 and 3.8 for [$\overline{1}10$] and [001] directions at 5 K, respectively. These values are comparable to reported results of CePd$_3$ (3.07)~\cite{webb1986observation}, Nd$_{0.95}$TiO$_3$ (3.63)~\cite{yang2006temperature}, and URu$_2$Si$_2$ (4)~\cite{nagel2012optical}.  Therefore, although it has been suggested that the deviations from Gurzhi scaling may be due to the presence of elastic but energy-dependent scattering, we demonstrate here that another possibility is that multiple weakly frequency-dependent conductivity channels contribute to the transport.


The presence of two conducting channels which turn into one as interband scattering increases means a violation of Matthiessen's rule in which one adds scattering rates for different kinds of scattering events that are independent of each other. In fact, a negative deviation from Matthiessen's rule was observed for SrRuO$_3$ and CaRuO$_3$ thin films~\cite{klein2001negative}. In that work, they introduced point defects by electron irradiation at low temperatures and found that the change in resistivity decreases with increasing temperature, in lieu of shifting the residual resistivity $\rho_0$ by a constant. According to Matthiessen's rule, the low-temperature resistivity for simple metals is $\rho = \rho_0 + A T^2$ in which $\rho_0$ is the residual resistivity and $A$ is a coefficient proportional to the density of states and the probability of two-body scattering. Usually both $\rho_0$ and $A$ are regarded as independent of temperature.    However, we would point out that the relative importance of the coupling of bands may changed with increased impurity scattering, which will also lead to deviations from Matthiessen's rule.


Finally, we address the consistency of our data with the older interpretation where a fractional power-law form was found~\cite{dodge2000low}.  In the present case, the availability of high-quality thin films and the development of THz spectroscopy has allowed us to resolve that the true low temperature form of the complex conductivity is two Drude terms. However, the fractional power-law form (Eq. \ref{frac}) used previously for SrRuO$_3$~\cite{dodge2000low} is still consistent with our data albeit over a restricted frequency range.  Over a restricted frequency range, the fractional power (e.g., $\alpha =0.5$) can be written as a sum of two terms, a Drude term in which $\alpha = 1$ and a constant (which is the approximate form of a Drude terms whose scattering rate is much larger than the measured frequency range). \begin{equation}
\begin{split}
	\sigma(\omega) =  \frac{\sigma_1}{(1-i \omega/\gamma_1)^1} +\frac{\sigma_2}{(1-i \omega/\gamma_2)^0}  \\
	\approx  \frac{\sigma_3}{(1-i\omega/\gamma_3)^{\alpha}}.
\end{split}
\end{equation}

\section{Conclusions}

In conclusion, we have shown that the optical conductivity of three metallic ruthenates, SrRuO$_3$, CaRuO$_3$, and Sr$_2$RuO$_4$, can be phenomenologically modeled with two Drude terms with different scattering rates at low temperatures. With increasing temperature, we find that there is a crossover from two-Drude to single-Drude conductivity.  Although details of the materials matter, this may indicate there is a general mechanism for non-Drude and non-universal low-energy optical conductivity in these moderately correlated 4\textit{d} transition metal oxides. 

The optical conductivity of  SrRuO$_3$ and Sr$_2$RuO$_4$ are qualitatively similar. Below a temperature scale of 30 K, the scattering rates of both Drude peaks can be roughly fit with $T^2$, which is expected for two-particle scattering. Owing to the presence of the narrow Drude peak whose scattering rate is 0.2 THz at 5 K, the renormalized scattering rate $\gamma(T, \omega)$ does not follow $\omega^2$ scattering rate in this frequency range. The low-temperature conductivity of CaRuO$_3$ is separated into a narrow Drude peak and a broad constant incoherent background.  At elevated temperatures the two relaxation time scales merge into a single Drude peak, probably owing to enhanced interband scattering.    We believe that this behavior arises from multiple conduction channels that are weakly coupled at low temperatures, but more strongly coupled at higher temperatures.  This interpretation is consistent with the existence of multiple Fermi surfaces in these compounds and with the expected relative weakness of $\omega^2$ dependent scattering as compared to $T^2$ dependent scattering in the Fermi liquid treatment.  As we have discussed, the presence of multiple conduction channels with interband scattering instead of multiple scattering channels violates the underlying assumptions of Matthiessen's rule. In a multiband system where carriers have different masses and Fermi velocities, charge conduction can be separated into parallel channels. 

The important of Hund's coupling has been recently emphasized for multi-band transition metal oxides like the ruthenates~\cite{mravlje2011coherence,georges2013strong}.   Hund’s coupling is the energy associated with intra-atomic exchange, which lowers the interaction energy when two electrons are placed in different orbitals with parallel spin, as opposed to two electrons in the same orbital.   It provides a mechanism for large electron-electron correlations even in systems that are far from the Mott regime.  Except for near half-filling or a shell with a single electron or a single hole, a Hund's coupling gives two key effects that compete with each other.  First, it tends to increases the critical U above which a Mott insulator is formed hence pushes a system away from the Mott regime.  On the other hand, it also tends to reduce the Fermi liquid coherence temperature scale and the energy scale below which a Fermi-liquid is formed, leading to a bad metallic regime in which quasiparticle coherence is suppressed.   This is an influence that has been called ``Janus faced".  The moderate scale of mass enhancements and the low Fermi liquid coherence temperature in the ruthenanes have been taken to be a signature of the Hund's coupling~\cite{mravlje2011coherence,georges2013strong}.  With regards to our present findings, although we have taken interband coupling to be of a generic variety, it may be that these effects of it at higher temperature and its suppression at lower temperature are both enhanced by these Hund's effects~\cite{georges2013strong} that can reduce the effective interaction in different orbitals~\cite{fanfarillo2015electronic}. In this regard, the very different scattering rates of the channels at low temperatures may be understood as consequence of the ``orbital-decoupling" effect of the Hund's effect.   Moreover,  the crossover to a single Drude term at higher temperatures can occur to an enhanced interband electron scattering that takes place above the Fermi liquid coherence temperature.  Recent optical work on Sr$_2$RuO$_4$~\cite{stricker2014optical} has shown the existence of excess spectral weight above 0.1 eV that was shown to be the consequence of the interplay between effects which allowed residual quasiparticle-like excitations at high energies and Hund's coupling.    This assignment was made via a comparison of the optical conductivity and DFT+DMFT calculations.  It would be interesting to apply such analysis to the current case where multi-component features are found in the optical conductivity at even lower energies.

We believe the experimental findings in this paper are general to the ruthenate family. However, they may shed light on low-disorder multiband metals in general and the anomalous transport behavior of other strongly correlated metals with complicated Fermi surfaces. Our result might be an example of correlated materials, for which the low-energy and low-temperature electrodynamics can be still grossly understood in terms of Fermi-liquid notions and semi-classical equations of motion. Nevertheless, it should be emphasized that our measurement is in the long wavelength limit and is therefore more sensitive to the collective behavior of the entire Fermi surface and almost conserved quantities.  They are not necessarily inconsistent with previous measurements of more disordered thin films or measurements in the near-infrared frequencies, which may be more sensitive to collective excitations and short wavelength correlations.

\begin{acknowledgements}
We would like to thank D. Maslov for helpful conversations.  Work at J.H.U. was supported though the National Science Foundation (NSF) Grant No. DMR-1905519.   Research at Cornell was supported by the National Science Foundation (Platform for the Accelerated Realization, Analysis and Discovery of Interface Materials (PARADIM)) under Cooperative Agreement No. DMR-1539918. N.J.S. acknowledges support from the NSF Graduate Research Fellowship Program (GRFP) under Grant No. DGE-1650441. Work by D.E.S.,  J.P.R., and K.M.S. was supported by the NSF through Grant No. DMR-1709255. This research is funded in part by the Gordon and Betty Moore Foundation (GBMF)’s Emergent Phenomena in Quantum Systems (EPiQS) Initiative through Grant Nos. GBMF3850 and GBMF9073 to Cornell University. This work made use of the Cornell Center for Materials Research (CCMR) Shared Facilities, which are supported through the NSF MRSEC Program (No. DMR-1719875). Substrate preparation was performed in part at the Cornell NanoScale Facility, a member of the National Nanotechnology Coordinated Infrastructure (NNCI), which is supported by the NSF (Grant No. ECCS-1542081). 
\end{acknowledgements}


\end{document}